\documentstyle[12pt]{article}

\newcommand{\ba}{\begin{eqnarray}}
\newcommand{\ea}{\end{eqnarray}}
\newcommand{\be}{\begin{equation}}
\newcommand{\ee}{\end{equation}}
\newcommand{\bib}{\bibitem}
\newcommand{\ed}{\end{document}}
\newcommand{\nn}{\nonumber\\}
\newcommand{\fr}{\frac}

\begin{document}
%\color{red}

\thispagestyle{empty}
%\begin{raggedleft}
%IF-UFRJ/xx/97\\
%hep-th/9707204\\
%July/97\\
%\end{raggedleft}
$\phantom{x}$\vskip 0.618cm

%\vfill

%\vfill

\begin{center}
{\huge On the dimensional dependence of duality groups for massive p-forms}\\
\vspace{1.5cm}
{\Large  J.L. Noronha, D. Rocha, M.S. Guimar\~aes, and C. Wotzasek}\\
\vspace{1.5cm}
{\em Instituto de F\'\i sica\\Universidade
Federal do Rio de Janeiro\\21945, Rio de Janeiro, Brazil\\}
\end{center}

\begin{abstract}
\noindent 
We study the soldering formalism in the context of abelian p-form theories.  We develop further the fusion process of massless antisymmetric tensors of different ranks into a massive p-form and establish its duality properties.
To illustrate the formalism we consider two situations.  First the soldering mass generation mechanism is compared with the Higgs and Julia-Toulouse mechanisms for mass generation due to condensation of electric and magnetic topological defects.  We show that the soldering mechanism interpolates between them for even dimensional spacetimes, in this way confirming the Higgs/Julia-Toulouse duality proposed by Quevedo and Trugenberger \cite{QT} a few years ago. Next, soldering is applied to the study of duality group classification of the massive forms.  We show a dichotomy controlled by the parity of the operator defining the symplectic structure of the theory and find their explicit actions.
\end{abstract}

\newpage

The study of duality transformation has had strong impact over different areas of physics -- from Strings to Condensed Matter -- with particular emphasis in the massless p-form electromagnetic theory.
Indeed duality operation and self-dual actions were thoroughly studied in the context of massless four dimensional electromagnetic theory and its even dimensional p-form extensions \cite{massless1,massless2}.
However a similar study in the context of massive abelian p-forms has attracted little interest \cite{QT,MCD} which prompts us to present this investigation.  We bring about a study that extends both the notion of duality symmetry to massive totally antisymmetric tensors of arbitrary ranks and the notion of self-duality.  Consequences of the present analysis to topological mass generation and duality group classification will be presented.

The ubiquitous role of the duality operation in the investigation of concrete
physical systems is by now well recognized \cite{reviews}.  This is a symmetry
transformation that is fundamental for investigations in arenas as distinct as
quantum field theory, statistical mechanics and string theory.
Establishing a duality means that one has  two equivalent mathematical description
of the same physical phenomenon in terms of different fields.
Duality is a general concept relating physical quantities in different regions of
the parameter space.
It relates a model in a strong coupling regime to its dual version working in a
weak coupling regime, providing valuable information in the study of strongly
interacting models.  The instance of electromagnetic-like self duality present in $D=4k+2$ dimensions has attracted
much attention because it seems to play an important role in many theoretical
models \cite{EK}.

It is important to mention that antisymmetric tensor gauge theories have attracted much interest in constructing 
gauge theories of elementary extended  objects (strings, membranes,...)
in  recent years \cite{ast,PO}. An antisymmetric tensor of rank $(p+1)$\ couples to elementary 
$p$-branes, a natural generalization of the coupling of the
vector potential one-form in Maxwell theory 
to elementary point-particles (0-branes). 
They also appear naturally
in effective field theories for the low-energy dynamics of strings where they play an important role in the realization of various dualities among different theories \cite{pol}. 
The study of dualities is becoming more and more important due to recent
developments in string theory, where it was shown that inequivalent vacua are 
related by dualities based on the existence of extended objects,
the D-branes \cite{rev}.

For massless p-forms duality group classification and mass generation mechanisms have not been thoroughly studied. Indeed the dependence of duality with dimensionality appears to be crucial.  The current status is as follows. The distinction among different dimensions is manifest by the double
dualization operation following from the identities, valid in Minkowskian spacetimes
\begin{equation}
\label{10}
\mbox{}^{**}F = \cases{+F,&if $D=4k+2$\cr
	-F,&if $D=4k$\cr}
\end{equation}
where $*$ denotes the usual Hodge operation and $F$ is a $\frac D2$-form.  This leads to two, apparently independent, consequences.  First, the concept of self-duality seems to be well defined only in twice odd dimensions, and not present in the twice even cases.  Second, (\ref{10}) leads to separate outcomes regarding the duality groups. 
The invariance of the action in different
$D$-dimensions is preserved by the following groups,
\begin{equation}
\label{20}
{\cal G}_d= \cases{Z_2,&if $D=4k+2$\cr
	SO(2),&if $D=4k$\cr}
\end{equation}
which were coined as ``duality groups". The duality operation is characterized by an one-parameter SO(2) group of symmetry in D=4k dimensions, while for D=4k+2 dimensions it is manifest by a discrete $Z_2$ operation. The $Z_2$ is a discrete
group with two elements.
This led to the prejudice that only the 4 dimensional Maxwell theory and its 4k extensions would possess duality as a symmetry, while for the 2 dimensional scalar theory and its 4k+2 extensions duality is not even definable. 
This two-fold dependence of duality with dimensionality has been clarified by algebraic methods \cite{massless1}, and its physical origin disclosed by the soldering \cite{massless2}.  The solution for these problems came, first with the recognition of a 2-dimensional internal structure hidden in the space of potentials \cite{Zwanziger,massless1,massless2}.  Transformations in this internal duality space have extended the concept of self duality to all even dimensions.  In the D=4 the explicitly self-dual Maxwell theory is known under the names of Schwarz and Sen but this deep unifying concept has also been appreciated by others \cite{massless1, Zwanziger}.

The investigation of mass generation for compact antisymmetric tensors of arbitrary ranks in $D=d+1$ dimensions, coupled to magnetic and electric topological defects, due to some condensation mechanism has been tackled by Quevedo and Trugenberger \cite{QT} that also established an interesting duality between the Higgs and the Julia-Toulouse mechanisms for even dimensional spacetime. In compact antisymmetric field theories p-branes appear as topological defects of the original theory. While electric
(p-1)-branes coupled minimally with the original p-forms, the magnetic (d-p)-branes can be viewed as closed singularities (Dirac strings).  The effective, low-energy field theory, is then valid outside these singularities. It is known that topological defects condensation leads to drastic modifications of the infrared behavior of the original theory \cite{PO,poly}. There is a new phase with a continous distribution of topological defects described by a low energy effective action - the condensation of topological defects gives rise to new low-energy modes representing the long-wavelength fluctuations about the homogeneous condensate. Quevedo and Trugenberger have shown that, in the presence of a magnetic defect described by a Dirac string $\psi^{(0)}_{p}$, a massless abelian $(p-1)$-form $\phi_{p-1}^{(0)}$ interpolates into a massive $(p)$-form $\psi^{(m)}_{p}$ in the condensed phase of the magnetic defect. In this process, coined by Quevedo and Trugenberger as Julia-Toulouse mechanism, the degrees of freedom of the abelian $(p-1)$-form are incorporated by the magnetic condensate to acquire a mass proportional to the density of the condensate,
\be
\phi_{p-1}^{(0)} \to  \psi_{p}^{(m)}=
\phi_{p-1}^{(0)} \oplus \psi_{p}^{(0)} \ee
This is quite distinct from the Higgs mechanism where the original $U(1)$ massless tensor $\phi^{(0)}_p$ acquires the degrees of freedom of the Higgs condensate, say $\Sigma_{p-1}^{(0)}$ to become massive,
\be
\phi^{(0)}_p  \to  \phi^{(m)}_p
= \phi^{(0)}_p \oplus \Sigma_{p-1}^{(0)} \ee
When the topological defects have the same dimensionality, Higgs and Julia-Toulouose phases are described by tensors of the same rank this way establishing a duality between these two mechanisms \cite{QT}.

Alternatively, one of us and collaborators \cite{massless2}, have developed a systematic method for the study of mass generation and different aspects of duality, that embraces all ranks and dimensions.  It contains two basic elements; a point contact transformation and the soldering formalism that operates in the internal space \cite{soldering}.
The mass generation is a consequence of the fusion of distinct fields and has been established in some special examples \cite{abreu}.
The soldering may be summarized by the following scheme,
\be
A^{(0)}_p  \to  A^{(m)}_p
 = A^{(0)}_p \oplus B_{q}^{(0)} \ee
if the ranks $p$ and $q$ of the massless fields $A_p^{(0)}$ and $B_q^{(0)}$ satisfy a massive duality condition: $p+q=d$.

In another direction, the ability of the formalism to distinguish among different group structures of the duality transformation is traceable to the dimensional dependence of the parity property of a differential operator (a generalized {\it non-covariant} curl) that solves the Gauss law of the massless components.  The parity's dimensional dependence  of the operator selects the proper symmetry in the symplectic sector of the theory while the Hamiltonian, being quadratically dependent on the operator does not differentiate between the two cases.
It is crucial to note that this procedure produces two distinct classes of dual theories characterized by the opposite signatures of the (2x2) matrices in the internal space.  These actions correspond to self-dual and anti-selfdual representations of the original theory. An interesting duality between these duality symmetric actions has been reported recently \cite{DS}.

In this work we study the soldering of arbitrary massless anti-symmetric tensors leading to a new, massive anti-symmetric form.  The duality transformation properties for this massive form is then studied from this point of view.  We show that for massive p-forms in odd-dimensional spacetime there is a dichotomy that very much resemble the analogous situation of the case of massless p-forms in even dimensional spacetimes.
The duality groups are either $Z_2$ or $SO(2)$  according to the dimension of the spacetime being $D=3$ mod(4) or $D=5$ mod(4) respectively. 
For the latter there is an one-parameter continuous $SO(2)$ invariance for the potentials in the internal space while for the former there is a $Z_2$ property that is not implementable as a canonical transformation but possess a self-dual form that is absent from other case.
We show that for massive p-form case it is the parity dependence on dimensionality of a ({\it covariant}) generalized curl that selects between an $SO(2)$ or a $Z_2$ structure for the theory.  However, the physical origin of the operator carrying this property is different since there is no Gauss law to restrict the space of states.  In the analysis that follows we use the {\it covariant} version of the dual-projection technique developed in \cite{massless2}, that has proved to be adequate to disclose duality and self-duality in the context Maxwell-like massless $p$-forms, to deal with the duality group dimensional dependence problem for massive p-forms.  This technique is based on the doubling of the space of potentials in order to make duality an explicit symmetry by creating an internal space of potentials.  Unlike the
algebraic or group theoretical methods, this technique is applicable to both even and odd dimensions as well as massless and massive $p$-forms therefore allowing this investigation.

Let us consider a massive p-form theory (basically a D-dimensional Proca-like model) whose action is defined as,
\be
\label{P10}
S_p^m = \left\langle \fr{(-1)^p(D-p-1)!}{2\;(p+1)!} H_{p+1}^2(A_p) + (-1)^{p+1} \fr{m^2}2 A_p^2\right\rangle\; ,
\ee
where $\langle\cdots\rangle$ means spacetime integration and we adopt a simplified notation that goes as follows - $A_p \equiv A_{\mu_1 \cdots\mu_p}$ represents a p-form and
\be
\label{P20}
H_{p+1}(A_p) = \partial_{[\mu_1}A_{\mu_2\cdots\mu_{p+1}]}\, ,
\ee
denotes its field tensor.  We use the following metric $g^{\mu\nu}=diag(+,-,\cdots ,-)$. After the elimination of the redundant variables using the constraints this model displays a total of $D-1\choose p$ degrees of freedom.
Making use of the identity
\be\label{P30}
\epsilon_{pq}\;\epsilon^{p\tilde q} = (-1)^{D+1}\; p! \;\delta_{[q]}^{\;\tilde q}
\ee
with
\be
\delta_{[q]}^{\;\tilde q} \equiv \delta_{[\mu_1}^{\;\tilde\mu_1}
\cdots\delta_{\mu_q]}^{\tilde\mu_q}\, ,
\ee
the action for the massive theory may be rewritten in a first-order form with the introduction of an auxiliary field
\be\label{P40}
S_p^m = \left\langle\Pi_q\;\epsilon \partial A_p - \fr{(-1)^q}2 \Pi_q^2 +\;(-1)^{p+1}\; \fr{m^2}2 A_p^2\right\rangle\, .
\ee
We shall denote the first term in this action as the ``covariant symplectic sector" and the remaining of the action as the symplectic potential or just potential for short \cite{fadjack}.
Here $\Pi_q$ is a q-form auxiliary field and
\be
\Pi_q\;\epsilon \partial A_p = \epsilon^{\mu_1\cdots\mu_q\alpha\mu_1\cdots\mu_p}
\Pi_{\mu_1\cdots\mu_q}\partial_\alpha A_{\nu_1\cdots\nu_p}\, .
\ee
This brings naturally the generalized-curl operator ($\epsilon\partial$) into the ({\it covariant}) symplectic sector of the action.  This is unlike the case of massless p-forms where the ({\it non-covariant}) curl operator was brought into the symplectic part of the action as the result of the solution of the Gauss law.
A further relabeling as
\be
\Pi_q = m B_q
\ee
and a simultaneous duality transformation of both $A_p$ and $B_q$ \cite{RJ-JP} will rephrase the action as 
\begin{equation}
\label{BI40}
S_{eff} = \left\langle\left(-1\right)^{p} \frac 18 \frac{p!}{(q+1)!} H^2_{q+1}(B_q)
                 +\left(-1\right)^{q} \frac 18 \frac{q!}{(p+1)!} F^2_{p+1}(A_p)
                 - m   B_{q} \epsilon \partial A_{p}\right\rangle\; ,
\end{equation}
which we recognize as a $B\wedge F$ type theory.

It is interesting, at this juncture, to compare the field contents of the soldering analysis with the mass generation coming from the Higgs and the Julia-Toulouse mechanism.  By inspection, we see,
\begin{itemize}
\item {Higgs/Soldering}
\be
\Sigma_{p-1}^{(0)} = \mbox{}^*\left(B_q\right)
\ee
\item{Julia-Toulouse/Soldering}
\be
\phi_{p-1}^{(0)} = \mbox{}^*\left(B_q\right)
\ee
\end{itemize}
where $*$ here is the massless duality operation, characterized by 
\be
\alpha_p = \mbox{}^*\beta_q
\ee
if $p+q=d-1$.  Therefore, in order to identify the fields we need the condition, $p-1 = q = d-p$ or, equivalently, $2p = d+1 = D$, that is the Quevedo-Trugenberger condition for the Higgs/Julia-Toulouse duality, to hold.  The field $B_q$ therefore interpolates between the original abelian form in the Julia-Toulouse condensation to the Higgs condensate in the Higgs mechanism.

Next, let us consider the classification of the duality group structure induced by the soldering analysis. Notice that the condition $p=q$ must hold for self-duality be manifest. This establishes the connection between the tensorial rank and spacetime dimensions
\be
D=2p+1\, ,
\ee
so that only odd dimensional manifolds are prone to display self-duality.  To recognize, among the odd dimensions, those displaying the $Z_2$ structure characteristic of the self-duality and those carrying the $SO(2)$ invariance, a field redefinition that rearranges the ({\it covariant}) symplectic sector of the action (\ref{P40}) is implemented.  Therefore, let us write
\ba\label{P50}
A_p &=& \left(A_p^+ + A_p^-\right)\nn
\Pi_p &=& \eta\; m \left(A_p^+ -A_p^-\right)\; ,
\ea
with $\eta=\pm$ being the signature of the canonical transformation, and bring this redefinition back into the action (\ref{P40}) to obtain
\ba\label{P60}
S_p^m &=& \left\langle\eta\; m \left[\left(A_p^+\epsilon\partial A_p^+ - A_p^-\epsilon\partial A_p^-\right) + \left(A_p^+\epsilon\partial A_p^- - A_p^-\epsilon\partial A_p^+\right)\right]\right.\nn
&+& \left. (-1)^{p+1}\; m^2\left[\left(A_p^+\right)^2 + \left(A_p^-\right)^2\right]\right\rangle\, .
\ea
The parity of the generalized-curl is defined as
\ba\label{P70}
\left\langle A_p\;\epsilon\;\partial\;B_q \right\rangle= {\cal P}(\epsilon\partial) \left\langle B_q\;\epsilon\;\partial\; A_p\right\rangle
\ea
and its dependence on the rank of the p and q-forms, or dimensionality is given by
\ba\label{P80}
{\cal P}(\epsilon\partial)  &=& (-1)^{(p+1)(q+1)}\nn
&\vdots & \mbox{limit}\;\;q\to p\nn
&=& (-1)^{\fr{D+1}2}\, .
\ea
Therefore, for the self-dual case ($p=q$) we identify the first term in the symplectic sector of (\ref{P60}) as the one displaying the $Z_2$ aspect while the second term displays $SO(2)$ invariance.  The presence of both structures, simultaneously, in the action is however illusory. In those dimensions where one structure is clearly present the other is a total derivative and vice-versa, this being a consequence of the dependence of the operator's parity (\ref{P80}) on dimensionality.  It is clear then that the $Z_2$ structure survives whenever $D=3+4k$ but is absent for $D=5+4k$ when the $SO(2)$ structure takes over.  Therefore, for the first case, $D=3 \;\mbox{mod(4)}$
\ba\label{P90}
S_p^m \to S_p^+(A_p^+) + S_p^-(A_p^-)
\ea
where
\ba\label{P100}
S_p^\pm = \left\langle\pm\;\eta\; m A_p^\pm\epsilon\partial A_p^\pm - (-1)^p \; m^2(A_p^\pm)^2\right\rangle
\ea
while for the latter, $D=5$ mod(4), the action becomes
\ba\label{P110}
S_p^m \to \left\langle\eta\; m A_p^\alpha \varepsilon^{\alpha\beta}\epsilon\partial A_p^\beta - (-1)^p \; m^2 (A_p^\alpha)^2\right\rangle
\ea
where $A_p^\alpha = A_p^\pm$ and $\epsilon^{+-}=1$.  Due to the diagonalization imposed by the $Z_2$ structure each one of the actions in (\ref{P90}) carries one-half the total number of degrees of freedom but in the $SO(2)$ case there is no such a decomposition so that all degrees of freedom in the action in (\ref{P110}) are entangled out.  Observe that the self-dual structure inherent to (\ref{P90}) is obtained directly from (\ref{P40}) imposing the second-class constraint of self-duality
\ba\label{P120}
\Pi_p = \pm \; m \; A_p\, ,
\ea
therefore reducing to half the number of degrees of freedom.  This generalizes to $D=3$ mod(4) the self-dual construction of \cite{TPvN} which is known to be the massive analogous to the selfdual scalar proposed by Floreanini and Jackiw \cite{FJ} in D=2.  The $SO(2)$ invariant model in (\ref{P110}) resembles the model proposed by Schwarz and Sen in their study of duality for massless electromagnetic-like actions. As far as we know the massive action analogous of the Schwarz-Sen action has not been proposed before.  Although, because of the signature $\eta$ we have two possibilities for the action in (\ref{P110}) (in (\ref{P100}) $\eta$ has no consequences) this should not be confused with the two chiral-like projections obtained from (\ref{P120}) leading to the two actions in (\ref{P100}).   Indeed, while for $D=3+4k$ the symplectic sector breaks into two disconnected pieces, each carrying half the number of degrees of freedom, the theory living in $D=5+4k$ is just reorganized into an $SO(2)$ like structure but there is no reduction of the ``covariant phase space".

In summary, we have studied the soldering of a massless p-form $A_p^{(0)}$ with an also massless q-form $B_q^{(0)}$ satisfying the massive duality condition $p+q=d$.  We have shown that as a consequence of soldering, a massive anti-symmetric tensor appears.  Comparison of this mass generation mechanism with both Higgs and Julia-Toulouse condensation approaches shows that it interpolates between them if the Quevedo-Trugenberger condition is satisfied.  This gives an alternative point of view in the Higgs/Confinement duality and sheds light in the meaning of the soldering mass generation.

With regard to the duality group classification, we have studied the structure of the duality groups for massive p-forms and disclosed its dimensional dependence.  We found that it is the dimensional dependence of the parity property of a generalized curl-operator, used in the reduction of the action to first-order, the responsible for the selection of the proper sector of the {\it covariant} symplectic part of the action. This is the sector presenting dimensional dependence since the potential, being quadratically dependent on the curl-operator is not sensible to this property.
A similar situation also happens for the massless case. There however it was the parity's dependence on dimensionality of the {\it non-covariant} generalized curl, coming from the resolution of the Gauss constraint, the property selecting the group commanding sector of the action.

We found that for $D=5$ mod(4) the theory is $SO(2)$ invariant while for $D=3$ mod(4) it is symmetric under the interchange of its self and anti-self-dual components but it has no generators.
This characterization of the massive modes in odd dimensions completes the table of duality transformations for electromagnetic-like p-forms.  It displays an intriguing sequence of doublets of $Z_2$ and $SO(2)$ for massless and massive theories in alternating even and odd dimensions starting with the former, i.e., $Z_2$ for $D=2,3$ and $SO(2)$ for $D=4,5$ and so on. These features are illustrate in the following table, 
\begin{center}
\begin{tabular}{|c|c|c|c|c|}
\hline
Dimensions  & massive &  massless  &  rank  & degrees of freedom     \\ \hline
13 &  SO(2)   &         &  6  & 924   \\ \hline
12 &          &  SO(2)  &  5  &  252    \\ \hline
11 &  $Z_{2}$   &         &  5  &  252    \\ \hline
10 &          &  $Z_{2}$  &  4  &   70    \\ \hline
9  &  SO(2)   &         &  4  &   70    \\ \hline
8  &          &  SO(2)  &  3  &   20    \\ \hline
7  &  $Z_{2}$   &         &  3  &   20    \\ \hline
6  &          &  $Z_{2}$  &  2  &   6     \\ \hline
5  &  SO(2)   &         &  2  &   6     \\ \hline
4  &          &  SO(2)  &  1  &   2     \\ \hline
3  &  $Z_{2}$   &         &  1  &   2     \\ \hline
2  &          &  $Z_{2}$  &  0  &   1     \\ \hline
\end{tabular}
\end{center}
The connection, by dimensional reduction, preserving the symmetry group, between the SD model in $D=3$ and the Schwinger model in $D=2$ is well known \cite{Govindarajan:1985ff}.
We have also found a non-preserving symmetry connection, by dimensional reduction, from $D=6$ to $D=4$ and from $D=4$ to $D=2$ in the massless case and from $D=5$ to $D=3$ in the massive case very much in the line of \cite{rdim} and hope to extend this connection to all the dimensions and symmetry groups in a future work through dimensional reduction technique \cite{future}.

Finally, as the result of the analysis developed in the present report, we found new actions for massive p-forms carrying explicitly the $Z_2$ and $SO(2)$ symmetries.  For the first case we found all the dimensional extensions of the $D=3$ action proposed by Townsend, Pilch and van Nieuwenhuizen \cite{TPvN} while the actions found here to carry $SO(2)$ symmetry, in any dimension, are new results.\\

{\bf Acknowledgments}  This work is partially supported by CNPq, FUJB, CAPES, PROCAD and FAPERJ,
Brazilian scientific agencies.

\ed

To The Editor,
Physics Letter B
Professor M. Cvetic
Dept. of Physics and Astronomy
University of Philadelphia
Philadelphia, PA 19106, USA
plb@cvetic.hep.upenn.edu

Dear Editor,

we are submitting our paper entitled "On the dimensional dependence of duality groups for massive p-forms" to be considered for publication in Physics Letter B.

                              Thanking you,
                            Yours sincerely,
                              C.Wotzasek
                       e-mail: clovis@if.ufrj.br